\documentclass{ws-procs9x6}

\begin{document}

\newcommand\vev[1]{\langle{#1}\rangle}

\def\la{\mathrel{\mathpalette\fun <}}
\def\ga{\mathrelbe {\mathpalSnon-gette\fun >}}
\def\fun#1#2{\lower3.6pt\vbox{\baselineskip0pt\lineskip.9pt
        \ialign{$\mathsurround=0pt#1\hfill##\hfil$\crcr#2\crcr\sim\crcr}}}

\renewcommand\({\left(}
\renewcommand\){\right)}
\renewcommand\[{\left[}
\renewcommand\]{\right]}

\newcommand\del{{\mbox {\boldmath $\nabla$}}}

\newcommand\eq[1]{Eq.~(\ref{#1})}
\newcommand\eqs[2]{Eqs.~(\ref{#1}) and (\ref{#2})}
\newcommand\eqss[3]{Eqs.~(\ref{#1}), (\ref{#2}) and (\ref{#3})}
\newcommand\eqsss[4]{Eqs.~(\ref{#1}), (\ref{#2}), (\ref{#3})
and (\ref{#4})}
\newcommand\eqssss[5]{Eqs.~(\ref{#1}), (\ref{#2}), (\ref{#3}),
(\ref{#4}) and (\ref{#5})}
\newcommand\eqst[2]{Eqs.~(\ref{#1})--(\ref{#2})}

\newcommand\pa{\partial}
\newcommand\pdif[2]{\frac{\pa #1}{\pa #2}}
                        

\newcommand\yr{\,\mbox{yr}}
\newcommand\sunit{\,\mbox{s}}
\newcommand\munit{\,\mbox{m}}
\newcommand\wunit{\,\mbox{W}}
\newcommand\Kunit{\,\mbox{K}}
\newcommand\muK{\,\mu\mbox{K}}

\newcommand\metres{\,\mbox{metres}}
\newcommand\mm{\,\mbox{mm}}
\newcommand\cm{\,\mbox{cm}}
\newcommand\km{\,\mbox{km}}
\newcommand\kg{\,\mbox{kg}}
\newcommand\TeV{\,\mbox{TeV}}
\newcommand\GeV{\,\mbox{GeV}}
\newcommand\MeV{\,\mbox{MeV}}
\newcommand\keV{\,\mbox{keV}}
\newcommand\eV{\,\mbox{eV}}
\newcommand\Mpc{\,\mbox{Mpc}}

\newcommand\msun{M_\odot}
\newcommand\mpl{M_{\rm P}}
\newcommand\MPl{M_{\rm P}}
\newcommand\Mpl{M_{\rm P}}
\newcommand\mpltil{\widetilde M_{\rm P}}
\newcommand\mf{M_{\rm f}}
\newcommand\mc{M_{\rm c}}
\newcommand\mgut{M_{\rm GUT}}
\newcommand\mstr{M_{\rm str}}
\newcommand\mpsis{|m_\chi^2|}
\newcommand\etapsi{\eta_\chi}
\newcommand\luv{\Lambda_{\rm UV}}
\newcommand\lf{\Lambda_{\rm f}}

\newcommand\lsim{\mathrel{\rlap{\lower4pt\hbox{\hskip1pt$\sim$}}
    \raise1pt\hbox{$<$}}}
\newcommand\gsim{\mathrel{\rlap{\lower4pt\hbox{\hskip1pt$\sim$}}
    \raise1pt\hbox{$>$}}}

\newcommand\diff{\mbox d}

\def\dbibitem#1{\bibitem{#1}\hspace{1cm}#1\hspace{1cm}}
\newcommand{\dlabel}[1]{\label{#1} \ \ \ \ \ \ \ \ #1\ \ \ \ \ \ \ \ }
\def\dcite#1{[#1]}

\def\mathcalm{{\mathcal M}}
\def\mathcalp{{\mathcal P}}
\def\mathcalr{{\mathcal R}}
\def\mathcalpr{\mathcalp_\mathcalr}

\newcommand\bfa{{\bf a}}
\newcommand\bfb{{\bf b}}
\newcommand\bfc{{\bf c}}
\newcommand\bfd{{\bf d}}
\newcommand\bfe{{\bf e}}
\newcommand\bff{{\bf f}}
\newcommand\bfg{{\bf g}}
\newcommand\bfh{{\bf h}}
\newcommand\bfi{{\bf i}}
\newcommand\bfj{{\bf j}}
\newcommand\bfk{{\bf k}}
\newcommand\bfl{{\bf l}}
\newcommand\bfm{{\bf m}}
\newcommand\bfn{{\bf n}}
\newcommand\bfo{{\bf o}}
\newcommand\bfp{{\bf p}}
\newcommand\bfq{{\bf q}}
\newcommand\bfr{{\bf r}}
\newcommand\bfs{{\bf s}}
\newcommand\bft{{\bf t}}
\newcommand\bfu{{\bf u}}
\newcommand\bfv{{\bf v}}
\newcommand\bfw{{\bf w}}
\newcommand\bfx{{\bf x}}
\newcommand\bfy{{\bf y}}
\newcommand\bfz{{\bf z}}

\newcommand\sub[1]{_{\rm #1}}
\newcommand\su[1]{^{\rm #1}}

\newcommand\supk{^{(K) }}
\newcommand\supf{^{(f) }}
\newcommand\supw{^{(W) }}
\newcommand\Tr{{\rm Tr}\,}

\newcommand\msinf{M\sub{inf}}
\newcommand\phicob{\phi\sub{COBE}}
\newcommand\delmult{\Delta V_{\chi\tilde\chi{\rm f}}}
\newcommand\mgrav{m_{3/2}(t)}
\newcommand\mgravsq{m^2_{3/2}(t)}
\newcommand\mgravvac{m_{3/2}}

\newcommand\cpeak{\sqrt{\tilde C_{\rm peak}}}
\newcommand\cpeako{\sqrt{\tilde C_{\rm peak}^{(0)}}}
\newcommand\omb{\Omega\sub b}
\newcommand\ncobe{N\sub{COBE}}

\newcommand\sigtil{\widetilde\sigma_8}
\newcommand\gamtil{\widetilde\Gamma}

\def\tick{yes}


\title{Which is the best inflation model?}

\author{David H.\ Lyth} 

\address{Physics Department, Lancaster University, Lancaster LA1 4YB,  U.K.}

\maketitle

\abstracts{Reasonable-looking models of inflation are compared,
taking into account the possibility that the curvature perturbation might
originate from some `curvaton' field different from the inflaton.}

\section{Introduction}
After twenty years of trial and error, we are in posession of several
reasonable-looking models of inflation.
In this talk I have a stab at comparing the
pros and cons of these models. I should make it clear at the outset
that I am  looking
only for a model of  the last few tens of $e$-folds of inflation,
enough that the observable Universe starts out well inside the horizon.
This `observable' inflation 
takes place with $H$ at least five orders of magnitude
below the Planck scale, and wipes out practically all\footnote
{The  scalar fields, including the notorious dilaton, must obviously have
started out at  positions from which they can descend to their present
values. But there seems little reason to think that the 
selection by Nature of this initial condition is related to the selection 
of the model of observable inflation, any more than it is related to the
selection of the Standard Model.}
 memory of earlier epochs. Understanding it will be enough to be going on 
with!

The task of assessing a model of inflation has recently been complicated,
by the realisation that the curvature perturbation might not originate from
the vacuum fluctuation of the inflaton (the inflaton paradigm). 
Instead it might originate from 
the vacuum of some other `curvaton' field
\cite{lw,p03curv,variable,bck,bck2} (curvaton paradigm).
The inflaton paradigm \cite{book}
 strongly constrains the inflationary potential
through the  CMB normalization of the curvature
perturbation; also, with the advent of PLANCK, the inflaton paradigm will
rule out many models through their  prediction for the spectral index.
In contrast,  the 
curvaton paradigm imposes only weak constraints on the inflationary potential
\cite{dl}.

Nearly all inflation models invoke a slowly-rolling inflaton field
$\phi$, whose potential $V$ must satisfy the flatness conditions
$|V'|\ll V/\mpl$ and $|V''|\ll V/\mpl^2$. This has several implications
\cite{book}.
First,   the inflaton  mass 
 during inflation must satisfy
\begin{equation}
m^2 \ll V/\mpl^2 \simeq 3H^2
\dlabel{flatness}
\,.
\end{equation}
where $H$ is the Hubble parameter. Second, 
 the quartic self-coupling
$\lambda$ must very tiny. Third, to keep control of the infinite number
of  non-renormalizable terms that are generically expected in the potential,
inflation should presumably take place with  $\phi\ll \mpl$.
(As mentioned later, there is a proposal for eliminating such terms
so as to allow $\phi\gg \mpl$.)

To achieve the flatness,
supersymmetry is usually invoked with the  potential  generated by 
an $F$ term, $F\equiv \tilde M_S^2$, giving
\begin{equation}
V = \tilde M\sub S^4 - 3\mpl^2 \mgravsq
\dlabel{vfpot}
\,,
\end{equation}
where $\mgrav$ is the field-dependent gravitino mass evaluated
during inflation. 
However, the $F$ term  breaks supersymmetry which
typically generates a soft mass term at least of order 
$\tilde M_S^2/\mpl$.
In order to satisfy the 
 flatness condition  \eq{flatness}, the actual mass must be suppressed
below the generic value.
The suppression  would be  severe if
the terms of \eq{vfpot} canceled  during inflation, which should therefore
be avoided. Given that, the generic soft mass is  of order $H$, requiring
only mild suppression to achieve the flatness condition.

\section{Low-scale inflation}

I consider first `low-scale' inflation models, in which 
the scale  $\tilde M\sub S$
of SUSY breaking during inflation is the same as the scale $M\sub S$
of SUSY breaking in the vacuum. A typical value of $M\sub S$,
corresponding to gravity-mediated SUSY breaking in the vacuum,
is $M_S\sim  10^{10}\GeV$. The value 
 is  an order of magnitude or two
bigger for  anomaly-mediated SUSY breaking, and 
some orders of magnitude smaller for
 gauge- or gaugino-mediated SUSY breaking. 
The  lowest possible scale
 $M\sub S\sim \TeV$ is rarely invoked, because of the difficulty
of satisfying collider constraints within a reasonably simple framework.

{\bf Non-hybrid modular inflation}. The potential of  a modulus is usually 
supposed to come entirely from supersymmetry breaking,
 with a  potential  of
the form 
\begin{equation} V(\phi) =  M_S^4 f(\phi/\mpl)
\,, \label{vmod1}
\end{equation}
where  $M_S$ is the scale of supersymmetry breaking in the vacuum
and  $f \sim |f'| \sim |f''| \cdots \sim 1$. (In the case of gauge- or
gaugino mediation, where scales other than $M\sub S$ and $\mpl$ 
are relevant, it is not quite clear that this form should hold but it
seems to be  assumed in the literature.)

Expanding around the VEV we learn that the modulus
mass is of order $M_S^2/\mpl$, hence of order the  gravitino mass.
Expanding instead around a maximum, 
chosen as the origin of $\phi$, the potential
becomes
\begin{equation}
V(\phi) = V_0 - \frac12 m^2 \phi^2 + \cdots 
\label{vmod2}
\,,
\end{equation}
with again $m\sim M_S^2/\mpl$. But $V_0\sim M_S^4$, hence
$m\sim V_0^\frac12/\mpl$ which marginally violates the flatness
condition \eq{flatness}.
It might happen that the modulus potential is a bit flatter than
the above estimate 
so that \eq{flatness}  is satisfied. Otherwise
there will be
 `fast-roll' inflation 
\cite{bg,rsg}  which may not last for enough $e$-folds.

Even if it leads to slow roll, this type of
modular inflation cannot satisfy the CMB constraint, which 
 requires  \cite{book} 
$M_S \sim 10^{15}\GeV$. Two fixes have been proposed.
 Kadota and Stewart \cite{ks}   use 
the complex field $\Phi$ consisting of a  modulus and its axionic partner.
(They also have an alternative scheme using two complex moduli.)
The tree-level potential is  of the form \eq{vmod1}, approximated as
\begin{equation}
V = V_0 + m^2 \( -|\Phi|^2 + \frac1\mpl \( \Phi^3 
+\Phi^{*3} \) + \frac1{\mpl^2} |\Phi|^4 \)
\end{equation}
The  maximum of the  potential is supposed to be a  point of
enhanced symmetry (the fixed point of a symmetry group), so that the
1-loop  correction  can drive the maximum away from the origin to
form the rim of a crater. 
With  a finely-tuned choice of the 
initial condition, this two-field inflation model can
generate a  curvature perturbation with  the required magnitude
even with the generic (fast roll) mass.

Banks  \cite{banks99} instead identifies the  modulus 
with a bulk field in the Horava-Witten setup,
so that in \eq{vmod1}
$M\sub S$ is replaced by  $M^\frac32/\mpl^\frac12$ with $M\sim 10^{16}$ 
the GUT scale.
 If the potential is flat
enough for slow roll, the
 CMB normalization can then be satisfied. 
This is the only modular inflation model
which invokes a departure from \eq{vmod1}.

{\bf Hybrid modular inflation}.
In the models considered so far, the inflationary potential is a function
only of the inflaton field. The alternative
{hybrid inflation} paradigm \cite{linde91} supposes that the 
potential is a function of 
the inflation field $\phi$ {\em and} a `waterfall' field $\chi$,
with a tree-level potential 
\begin{equation}
V = V_0  -\frac12 m_\chi^2 \chi^2 + \lambda \chi^2 \phi^2 + \frac12 m^2\phi^2 
 + \cdots
\label{vhybrid}
\,.
\end{equation}
Inflation takes place in the regime $\phi>
\phi\sub c\equiv m_\chi/\sqrt{2\lambda}$, with $\chi=0$.
By `hybrid modular' inflation, I mean hybrid inflation in which 
(i) the waterfall field is a modulus and (ii) the inflaton mass
comes from gravity mediated supersymmetry breaking. Gravity
mediated supersymmetry breaking is also assumed for the MSSM,
so that {roughly} $V_0^\frac14\sim M\sub S\sim 10^{10}\GeV$
and $m\sim m_\chi \sim \mgravvac\sim 100\GeV$.
To be precise though, the model requires that the actual masses satisfy
 a  mild hierarchy $m\ll \mgravvac \ll m_\chi$.

In the original tree-level version \cite{rsg} this hierarchy is supposed
to be entirely accidental. However, the 
 maximum of the modulus
potential must represent a point of enhanced  symmetry, to 
ensure the absence of a linear term $\chi\phi^2$ \cite{clprep}.
It is therefore reasonable 
to suppose \cite{running} that radiative corrections lead to a running
mass $m(\phi)$.
With strong running,
 $m(\phi)$ can pass through
zero,  permitting slow-roll inflation even though its generic value
is of order $\mgravvac$.
Such strong running implies a strong running of the spectral index
(going in the opposite direction from the one proposed by the WMAP group)
which may rule out the model in the future \cite{clm}.

{\bf Inflation without slow roll}.
Two ways are known, by which the 
modular potential \eq{vmod1} can inflate without slow roll.
 {\em Thermal modular inflation} \cite{thermal2} invokes  a finite temperature 
contribution $\sim T^2\phi^2$, leading to inflation in the 
regime $m\lsim T\lsim M\sub S$ and $\mpl/M\sub S$ $e$-folds of inflation.
 {\em Locked inflation} \cite{dk} invokes 
 the same setup as hybrid modular inflation,
but with the alternative hierarchy $m\sim m_\chi\gg \mgravvac$;
then  $\phi$ oscillates until its amplitude is of order $\phi\sub c$,
leading to $\sim \frac32 \mpl/M\sub S$ $e$-folds of inflation.
By themselves, these schemes hardly  generate enough inflation,
but  extra fast-roll
inflation can occur later while the modulus descends to the vacuum
\cite{dl,ad},  and there may also be ordinary (late-time)
thermal inflation \cite{thermal1}.
More seriously, these schemes must be married to a curvaton scenario,
since they involve no slow-rolling inflaton. Still, it is interesting
that inflation can take place without any kind of slow roll.

{\bf The inflaton as a $\mu$ field}.
It is commonly supposed that the
 $\mu$ parameter of the MSSM is generated by
the VEV  of some  field which I will call the `$\mu$ field'.
This has led to two models of inflation. 
In the model of Dine and Riotto \cite{dr}
$\mu \sim \langle S^3\rangle /\mpl^2$ and  $|S|$ is the 
inflaton in a non-hybrid model. Gauge-mediated SUSY breaking is 
assumed for the MSSM,  with the inflaton potential  generated
by a tachyonic soft mass, and by terms $S^5/\mpl^2$ and $-XS^4/\mpl^2$ 
in the superpotential.  CMB normalization for this model requires
$V\ll \tilde M\sub S^4$ implying unacceptable fine-tuning of the soft 
mass. This would be avoided if the model could be
 married to some version of the curvaton paradigm.

In the model of 
King and collaborators \cite{bk,bck}, $\mu =\lambda \langle \mathcal N\rangle$
with $|\mathcal N|$
the waterfall field for a hybrid inflation. The mass of $\mathcal N$
is  $\kappa \langle S\rangle $ with $|S|$ the inflaton. The VEVs of
$\mathcal N$ and $S$ break Peccei-Quinn symmetry  so that the model contains
the QCD  axion.  CMB normalization for this model again requires
unacceptable fine tuning for the inflaton soft mass, which will
disappear if the model can be married to some version of the curvaton
paradigm. Preliminary results \cite{bck2} suggest 
that instead the Higgs flat direction may act as a curvaton, generating
the curvature perturbation through a preheating mechanism. 

\section{Other models using ordinary field theory}

To have an inflation scale  lower than $M\sub S$,
one would need
either a  cancellation of  the two terms of \eq{vfpot},
or a model in which the SUSY breaking scale $\tilde M\sub S$
is lower than $M\sub S$. Neither has been attempted, but there
are proposals for a higher inflation scale.

{\bf GUT inflation.}
Many models of inflation, starting with the original one \cite{new},
use a GUT Higgs field to generate a potential $V^\frac14 \sim 10^{15}\GeV$
or so. The archetypal proposal (with several variants) uses
 the superpotential \cite{cllsw}
$W=g S(
M^2-\Phi\bar\Phi)$, with $|S|$ the inflaton and $|\Phi|$ a Higgs field
which is the waterfall field, and with $M\sim 10^{16}\GeV$ the GUT scale.
This gives $\tilde M\sub S =\sqrt g M$. Assuming that the slope is dominated
by the 1-loop correction, the complete potential is \cite{dss}
\begin{equation}
V = g^2 M^4 \( 1 + c g^2 \ln \(\phi/Q\) \)
\label{vgut}
\,,
\end{equation}
with $c$ a loop suppression factor and $Q$ the renormalization scale.
For reasonable $c$, the CMB normalization is satisfied independently of
$g$. By  choosing  $g\sim 10^{-2}$ or so,
inflation takes place at $\phi\ll \mpl$, which may  justify
the neglect of non-renormalizable terms. Subject to the proviso that
GUT theories tend to be complicated, this type of model seems to be
quite attractive, though it does not explain why the inflaton mass
is significantly less than $H$.

{\bf $D$-term inflation.}
Inflation with a Fayet-Iliopoulos   $D$ term at the string scale
was  proposed,  in the context of the weakly coupled
heterotic string \cite{ewandtof,dterm}.
 Ignoring non-renormalizable terms it
gives the  potential \eq{vgut}, but now with
$g$  a gauge coupling and $M\sim 10^{17}\GeV$ the string scale.
However, $g$  presumably cannot now be  small, which means that
the inflaton field value of order $\mpl$ so that non-renormalizable
terms probably spoil the flatness \cite{km}.\footnote
{The  mass of order $H$
 is in contrast absent since we are not dealing with an $F$
term; this was the main motivation for the model.}
 Also, the high string scale gives the wrong CMB normalization.
Replacing the string scale by an ad hoc scale also leads to problems
\cite{accr2}. On the whole, I would say that $D$-term inflation was a 
promising idea that didn't  really work.

{\bf $D$--to--$F$ term inflation.} Instead of using the string-scale
FI $D$-term directly,  one may suppose that driving it to
zero  generates an $F$ term which can give inflation at a somewhat lower
scale.
It has been shown \cite{cllsw,ewandtof,glm}  how to construct models
of this type in the framework of the weakly coupled heterotic string.
The construction suppresses the mass of order $H$,
by making the inflaton the  pseudo-Nambu-Goldstone-boson (PNGB)
of a (rather complicated, stringy) symmetry.
However,   a working example has not yet  been 
written down, nor any connection made with a known (say GUT) extension of 
the Standard Model.

{\bf Models invoking an ordinary  global symmetry.}
One may also suppress the mass of order $H$ by using a PNGB associated with
 an ordinary  symmetry like $SU(N)$. One proposal \cite{ars}
invokes non-hybrid inflation,
with the inflaton mass canceled by the contribution of a PNGB, and another
\cite{cs} makes the inflaton itself a PNGB in a hybrid model. 
There has
not been any definite proposal for the origin of the inflation scale
in either of these schemes, and  they seem to be attractive only to the extent
that one takes seriously the mass of order $H$ problem. 
Much more
radically, it has been proposed to \cite{accr2} to use  global 
symmetry without supersymmetry. This type of inflation model is
 inspired by  `little higgs' extensions
of  the Standard Model, but have not been unified with one of them.
Such unification would require a high-energy completion of a 
`little higgs' model, since inflation presumably takes place far above
the scale $10\TeV$ of the `little hierarchy' which the little higgs takes care
of.

\section{Models going beyond ordinary field theory}

All  of the above models inflation (save the last)
 assume  some effective field theory, which is  constructed
along the same conservative lines as are customary when considering other
extensions of the Standard Model such as the MSSM, models of the axion or 
models for the origin of neutrino mass.
 Given that a  conservative
setup seems to work, it is not quite clear why one should look any further.
Still, I end by considering  some
 models which introduce radical new features. 

The oldest of them \cite{chaotic},  usually called
{\em chaotic inflation}, invokes a potential $V=\frac12m^2\phi^2$ or 
$V=\lambda \phi^4$, which is supposed to be valid right up to
 $\phi\sim 10 \mpl$ or so. This is indeed radical,
because one usually supposes that
 the tree-level potential will have
an  infinity of non-renormalizable terms  of order
$\phi^d/\mpl^{d-4}$. By virtue of the large field value \cite{mygrav},
 these models in contrast to the others give a significant tensor perturbation.
Because of that,  and the predicted tilt $n-1$, the quartic model
is in danger of being  ruled out by observation.

{\em  Gauge inflation}  \cite{accr1} makes the inflaton the fifth
component of a five-dimensional gauge field with Wilson Line symmetry
breaking.  {}The effective field theory of
 this proposal contains an infinite number of 
fields, a Kaluza-Klein tower. The
sinusoidal inflaton potential is obtained by 
summing the Coleman-Weinberg potentials of the entire tower, and there
has been much debate about the manner in which the sum converges
\cite{peter}. Leaving such worries aside, gauge inflation
 can justify the `chaotic' $\frac12m^2\phi^2$ proposal, though 
 according to \cite{banks03} there is no known 
 string-theoretic realisation.

{\em Colliding brane inflation}  \cite{dt}  identifies the inflation of 
 the 4-D effective theory, with 
with the distance between colliding $D$-branes in  extra dimensions.
The proposal has received enormous attention, 
because the calculation
of the potential within various setups apparently
involves deep issues of string theory. Taking these issues seriously,
it is not easy to get a flat  potential  which satisfies the CMB
constraint. But, as the  
 reheating process presumably cannot  be described within the same
effective 4-D theory as the inflation, it may be difficult to marry 
this type of inflation to a curvaton scenario which requires that
 the curvaton field exists  both during  and after inflation.

\end{document}